\documentclass[11pt,letterpaper]{article}
\usepackage{naaclhlt2012}
\usepackage{times}
\usepackage{latexsym}
\usepackage{amsmath}
\usepackage{multirow}
\usepackage{url}
\usepackage{epsfig}
\title{Graph-based Approach to Automatic Taxonomy Generation (GraBTax)}
\author{Pucktada Treeratpituk$^{\ddag}$, Madian Khabsa$^{\dag}$, C. Lee Giles$^{\ddag, \dag}$\\
  $^{\ddag}$Information Sciences and Technology, $^{\dag}$Computer Science and Engineering\\
  Pennsylvania State University, University Park, PA 16802, USA\\
  {\tt pucktada@gmail.com, madian@psu.edu, giles@ist.psu.edu}
}
\date{}

\begin{document}
\maketitle
\begin{abstract}

We propose a novel graph-based approach for constructing concept hierarchy from a large text corpus. Our algorithm, GraBTax, incorporates both statistical co-occurrences and lexical similarity in optimizing the structure of the taxonomy. To automatically generate topic-dependent taxonomies from a large text corpus, GraBTax first extracts topical terms and their relationships from the corpus. The algorithm then constructs a weighted graph representing topics and their associations. A graph partitioning algorithm is then used to recursively partition the topic graph into a taxonomy. For evaluation, we apply GraBTax to articles, primarily computer science, in the CiteSeerX digital library and search engine. The quality of the resulting concept hierarchy is assessed by both human judges and comparison with Wikipedia categories. 
\end{abstract}

\section{Introduction}

A taxonomy organizes concepts into a hierarchical structure, where broad concepts are at the top of the hierarchy and more specific concepts are further down. Large document collections are often organized into taxonomies, e.g. the Library of Congress and MEDLINE, because taxonomies enhance both search and browse features. In addition, a taxonomy also serves as a summary of a collection's content. Generally, taxonomies are manually created and maintained by domain experts, which is extremely time-consuming and costly. As a consequence, they are often incomplete and quickly become outdated, especially in a rapidly evolving domain like computer science. Thus, it is highly desirable to be able to generate taxonomies automatically. 

Some taxonomies have well-defined relationships between concepts; every child concept is related to its parent through the same relationship (either \textit{is-a} or \textit{part-of}). Others taxonomies are less rigid; concepts are connected to their parents with various types of relationship not necessary homogeneous. For instance, in ODP, the \textit{Outdoors} category contains \textit{Camping} (\textit{is-a}), \textit{Fishing} (\textit{is-a}) and \textit{Equipment} (\textit{tool-for}) as subcategories. The focus of this paper is on the second type of taxonomies. Even if such taxonomies are ``loosely'' defined, they are useful for browsing and visualizing a large corpus. 

There are two approaches to taxonomy generation: the query-independent approach, where one global taxonomy is constructed for the whole corpus \cite{mimno:hPAM07}, and the query-dependent approach, where a taxonomy is created for each query \cite{lawrie:hier_web03}. While a query-independent taxonomy offers a consistent view of the whole corpus, it makes an implicit assumption that there is one optimal taxonomy for that collection. A query-dependent approach, on the other hand, allows concepts to be organized differently depending on the query. Our approach present here focuses only on the query-dependent approach. 

Prior techniques used for automatic taxonomy generation can be grouped into three main categories: pattern-based, clustering-based, and knowledge source methods. Clustering-based methods hierarchically cluster topics based on similarity measures \cite{schmitz:www06,begelman:tag06,mika:web07}. The features used for calculating the similarity range from document vector, statistical co-occurrences to syntactic dependency. Pattern-based methods use lexico-syntactic patterns (such as ``\textit{NP, especially \{NP,\}*}'') to discover relationships between different concepts \cite{hearst:hyponym92,kozareva:hyponym08}. Knowledge source approach integrates information from existing knowledge sources, such as Wikipedia categories, ODP and WordNet, to identify the proper relationship between concepts \cite{van:folkson07,lin:onto09}. 

Pattern-based approaches are generally able to extract relations between concepts with high accuracy. However, such relations are limited to only those that explicitly appear in the corpus with those patterns. Therefore the coverage can be problematic, especially for specific concepts that appear infrequently. Similarly, knowledge source approaches can enhance the quality of the generated taxonomy, but are limited to domains where such resources exist. Clustering-based approaches, while in general do not have as high accuracy as the other two approaches, are the most flexible. They are able to discover relations, which may not explicitly appear in text. They also require minimum domain knowledge compared to the other two approaches. Our approach for taxonomy generation presented in this paper falls under the clustering-based methods.

Specifically, we propose a novel graph-based algorithm, called GraBTax (Graph-Based Taxonomy Generation), for automatically constructing a query-dependent taxonomy from a corpus. Our proposed algorithm has the following attractive characteristics. First, it incorporates both statistical co-occurrence and lexical similarity in determining the relationship between topics; the framework is also flexible enough to be extended to include other features. Second, the algorithm tries to construct a balanced taxonomy, where each topic is divided into distinct subtopics but of a similar generality level, though graph partitioning optimization. It also does not rely on any external knowledge sources. Thus, it could be easily applied to multiple domains. For evaluation, we apply the algorithm to generate taxonomies for topics in computer science using papers from CiteSeer$^X$ digital library.

\section{Related Work}
There has been significant interest in automatic taxonomy generation in Semantic Web, Information Retrieval and Knowledge Management communities \cite{cimiano:ontology06}. There are also efforts focusing, not on generation, but on extending existing taxonomies \cite{snow:semantic06,yang:metric09}.  

Much work has been done on organizing related user tags into clusters using tag co-occurrence \cite{schmitz:www06,begelman:tag06,mika:web07,wu:www06,specia:folk07}. Some used subsumption-based models to cluster concepts based on co-occurrence frequency \cite{sanderson:subsumption99,schmitz:www06}. Wu et al proposed a probabilistic model to generate groups of semantically related tags using the co-occurrence of tags, resources and users \cite{wu:www06}. Begelman et al represented tags as an undirected graph, where the weight on each edge is the co-occurrence frequency, and then used a spectral graph partitioning algorithm to generate hierarchical clusters of tags \cite{begelman:tag06}. This work is closely related to our approach. However, they mostly rely on user-generated tags and social networks of taggers, which are not available to GraBTax. They also only rely on co-occurrences, but not lexical similarity. Others also tried to enrich folksonomies with semantics by integrating other knowledge sources such as WordNet, Google and Semantic Web resources \cite{van:folkson07,lin:onto09}. But such resources are not always available. 

Many pattern-based approaches also have been proposed. Pattern-based methods define lexical-syntactic patterns for relations, such as is-a, part-of, and synonym, and use these patterns to discover instances of relations. These patterns can either be manually constructed \cite{berland:partof99,kozareva:hyponym08} or automatically bootstrapped \cite{hearst:hyponym92}. While pattern-based methods generally produce high accuracy relationships, they suffer from sparse coverage simply because many relationships cannot be found through pattern-matching. Other work used the nearly unlimited size of the web to solve the coverage problem \cite{etzioni:knowit05}. However, such methods have problems similar to those using resources like WordNet; when the domain of interest is very specific such as topics in computer science, most of the relationships cannot be found on the web through pattern-matching. 

Another related area of work is topic model research, which has recently become popular. \cite{blei:corr_topic07,griffiths:topic04,li:pachinko06} apply topic models to extract topics in scientific publications. Advanced topic models are able not only to identify topics, but also to discover relationships and organization between topics \cite{blei:corr_topic07,li:pachinko06}. For instance, the Pachinko Allocation Model (PAM) represents correlations between topics using a directed acyclic graph (DAG) \cite{li:pachinko06}. Its extension, the hierarchical PAM, explicitly generates a topic hierarchy \cite{mimno:hPAM07}. While the discovered topic word distributions are often meaningful, it is generally very difficult for a user to understand a topic just based on the resulting multinomial distribution \cite{mei:label07}.

\section{Graph-Based Taxonomy Generation}

The GraBTax algorithm can be decomposed into three main parts: topic extraction, graph construction and topic-dependent taxonomy generation. The first two steps are carried out offline, while the last step is carried out at query-time on a per-query basis. First, a set of candidate topics are automatically extracted from the corpus and the document co-occurrences between topics are calculated. Second, the algorithm constructs the topic associate graph encompassing all topics. Third, when a user issues a query topic, a topic-specific subgraph is selected and the graph partitioning algorithm is then used to convert the subgraph into a hierarchical taxonomy.
	
\subsection{Extracting Topics\label{sec:topic}}

Here, we describe how GraBTax derive the set of candidate topics from the corpus. Similar works in automatic taxonomy induction assume that terms in a taxonomy are given either from external sources such as Wikipedia category labels \cite{ponzetto:wiki07} or through user generated tags \cite{schmitz:www06,begelman:tag06}. GraBTax, on the other hand, does not rely on such external resources. Instead, the set of topics are automatically extracted from the corpus. 

Our algorithm makes one strong assumption regarding the nature of the data. It assumes that given enough number of documents, any meaningful topic will appear in multiple times in the document titles. And vice versa, if a term/phrase does not appear in any titles, then it does not warrant being included in a taxonomy. This is a reasonable assumption for scientific articles. By restricting the topic candidates to only terms appearing in the titles, we can easily obtain the set of candidate topics with high accuracy based on the most frequent variable-length ngrams appearing in the titles.

To construct the set of candidate topics, first, the algorithm separately generates word-level bigrams, trigrams and quadgrams that appear at least 3 times in titles of papers in CiteSeerX. Ngrams containing stopwords are ignored. Prepositions such as ``of'', ``for'', ``to'' are allowed to be present in the middle of ngrams. To take into account over-counting, the frequency of bigrams and trigrams are discounted. The list of bigrams, trigrams and fourgrams are then merged together to create one single list of ngrams. These are selected as candidate topics. For each topic pair, its co-occurrence within the document abstract is also computed.

\subsection{Constructing the Topic Association Graph}
Once we extract the set of topics and their co-occurrence, we then construct the topic association graph.
The topic association graph is defined as an undirected weighted graph, $G = (T, E)$, where both vertices and edges have weights. Each topic $t_i$ is a vertex in $G$ ($\forall_i\; t_i \in T$).  There exists an edge between the topic $t_i$ and the topic $t_j$ ($e_{ij} \in E$) if and only if $t_i$ co-occurs with $t_j$ in a document.
Now we precisely define how weights for each vertex and each edge are computed. These weights will be later used to determine which topics will be included in a taxonomy, and how to optimally partition a graph into a hierarchical taxonomy. Let the \textbf{strength} $s_i$ denote the weight for the vertex $t_{i}$, where $s_i$ is computed as:
\[s_i = \sum_{e_{ij} \in E} count(t_i, t_j)\]
where $count(t_i, t_j)$	is the number of documents that $t_i$ and $t_j$ co-occur in the title and document abstract. The strength of a topic, is simply the sum of all its co-occurrences representing that topic's importance. In addition to the strength measure, we also compute the degree for each vertex $t_i$, denoted by $k_i$, which is just the total number of edges associated with the vertex $t_i$.

Let $w_{ij}$ be the weight of the edge $e_{ij}$ between the topic $t_i$ and the topic $t_j$. The weight $w_{ij}$ depends not only on the co-occurrence between $t_i$ and $t_j$, but also on their lexical similarity. More precisely, an edge weight $w_{ij}$ is defined as: 

\begin{eqnarray*}
w_{ij} = [ 1 &+& \lambda_1  \: 1_{(rank(t_i|t_j) = 1 \:\text{ OR }\: rank(t_j|t_i) = 1)}  \\
			  &+& \lambda_2 \: jac(t_i,t_j) ] \times count(t_i, t_j) 
\end{eqnarray*}
	where
\begin{eqnarray*}
1_{cond}	&=& 1 \text{ if } cond \text{ is true, and } 0 \text{ otherwise } \\
rank(t_i|t_j)	&=& | \{\: t_h \:|\: s_j < s_h \text{ and }  \\
                &&   Pr(t_h|t_j) > Pr(t_i|t_j) \: \} | + 1 \\
jac(t_i, t_j)	&=& \text{Jaccard similarity between $t_i$ and $t_j$ }
\end{eqnarray*}
$rank(t_i|t_j)$, the rank of $t_i$ for $t_j$, is one plus the number of the topics with higher probability than $t_i$ given $t_j$, not counting the topics that have lower strength than $t_j$. If $t_i$ has the highest conditional probability given $t_j$, then the rank of $t_i$ for $t_j$ is 1. For example, for topic $t_j$ = ``\textit{vertex cover},'' the topic``\textit{approximation algorithm}'' and ``\textit{np complete}'' have the first and the second ranks respectively. The intuition is that if $t_i$ is highly predictive of $t_j$ (meaning if $t_i$ is present, it is  highly likely that $t_j$ is also present) or vice versa, then the strength of the connection between $t_i$ and $t_j$ should be higher than those indicated by their co-occurrence counts. Similarly, if topic $t_i$ and $t_j$ are similar lexically, as measured by the Jaccard similarity, then their connection receives a higher weight. 

Note that the weight $w_{ij}$ of the edge $e_{ij}$ is defined to incorporate both the statistical co-occurrence and the lexical similarity between topics $t_i$ and $t_j$. The relative weights to given each type of similarity ($\lambda_1$ and $\lambda_2$) are currently set heuristically.

\subsection{Selecting the Topic-Specific Subgraph}

Given a query topic $t_0$, first, a query-specific subgraph is selected from the topic association graph. This process determines which topics are to be included in the final taxonomy. 

Let $G_0 = (T_0,E_0)$ denote the topic-specific subgraph for the query $t_0$, the set of vertices ($T_0$) and edges ($E_0$) for the subgraph $G_0$ are defined as follow:
\begin{align*}
	T_0 = \{\; t_i \in T \;|\;	&rank(t_0|t_i) \leq r_{max} \text{ and } \\
						   		&k_i \geq k_{min}  \text { and } s_i \geq s_{min} \;\} \\
	E_0 = \{\; e_{ij} \in E \;|\;	&t_i, t_j \in T_0 \;\}					
\end{align*}

The threshold constant $k_{min}$ and $s_{min}$ denote the minimum degree and the minimum strength of topics to be selected respectively. They regulate the specificity of topics to be included in a taxonomy. Lowering either $k_{min}$ or $s_{min}$ allows the taxonomy to include more specific concepts. Similarly, by increasing them, only broader topics will be included, resulting in a smaller taxonomy. On the other hand, the threshold $r_{max}$ (the maximum rank) controls the relative-specificity with respect to the query topic $t_0$. With low value of $r_{max}$, only topics strongly related to the query topic will be included and vice versa. Also, note that the definition of $rank(t_0|t_i)$ permits that only the topics with lower strength than $t_0$ to be included in the subgraph for $t_0$.

If the query topic $t_0 \not\in T$ (e.g. $t_0$ is a unigram while $\forall t_h \in T$ is at least a bigram), then a dummy $t_0$ is created using a simple query expansion. For example, for $t_0$=``\textit{database}'', the topic-specific subgraph can be constructed based on the top three topics that are most similar to ``\textit{database}'', e.g. ``\textit{database system}'', ``\textit{relational database}'' and ``\textit{large database}.'' Thus, the topic-specific subgraph can be defined as $G_{database} = (T_{database},E_{database})$, where $T_{database} = T_{database\:system} \cup T_{relational\:database} \cup T_{large\;database}$.

\subsection{Partitioning the Topic-Specific Subgraph into a Taxonomy\label{sec:partition}}

Once the subgraph $G_0$ for the query topic $t_0$ is selected, the graph partitioning algorithm is applied to partition the subgraph into a taxonomy. First, all topics in the subgraph $G_0$, excluding the query topic, are divided into partitions. Within each partition, a topic vertex is selected to be the label of the partition. These label topics become the first-level subtopics of the query topic $t_0$ in the taxonomy. Then for each partition, all edges associated with its label topic are removed and the partition is further divided to generate the second-level subtopics. The partitioning is carried out recursively until a stopping criteria heuristic is met, which is when the number of topics in the partition is less than a minimum threshold or the intra-partition connectivity is zero. 

The number of partitions at each level is determined by the number of vertices in the parent partition. Let $n(G')$ denote the number of subpartitions to split a parent partition $G'=(T',E')$ into, then
\[ n(G') = \left\{ 
	\begin{array}{l l}
 		\lfloor (|T'|/ \beta) \rfloor 		& 	\quad \text{ if } |T'| < \alpha	\\
 		\alpha/\beta & 	\quad \text{ otherwise }	\\
	\end{array} \right. \]
where $\alpha$ and $\beta$ are constants. In our implementation, $\alpha = 200$ and $\beta = 20$.

A good taxonomy generally has two characteristics. First, each subtopic under the same parent should be quite distinct from each other. A taxonomy where sibling topics are very similar to each other is not very useful. Second, they should be of roughly equal generality/specificity level. Thus, when partitioning a parent topic into its subtopic partitions, there are two objectives to consider. First, we want to minimize the edge-cut of the partitioning. The edge-cut of a partitioning is the total weights of edges between topics belong to different partitions. Second, the sum of vertex's strength for each partition should be roughly equal. 

\begin{figure}
\centering
\epsfig{file=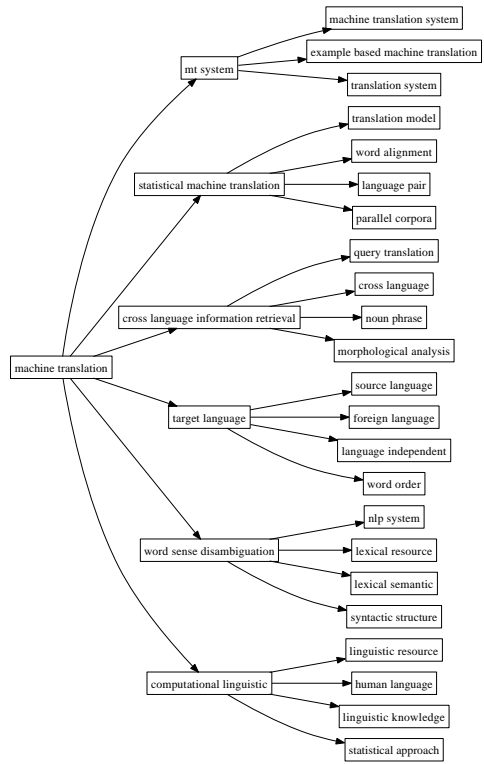, width=2.5in}
\caption{The two-level taxonomy for ``\textit{Machine Translation}'' as produced by GraBTax\label{fig:mt}}
\end{figure}

GraBTax uses the multi-level graph partitioning algorithm, proposed in \cite{karypis:irr_graph99,karypis:constraint98}, to find the optimal partitioning that minimizes the edge-cut while keeping the vertex strength balanced \cite{karypis:irr_graph99,karypis:constraint98}. The multi-level graph partitioning algorithm has been successfully used for load balancing in parallel computation environments, where multiple constraints such as memory load and storage requirement need to be balanced \cite{karypis:constraint98}, and for web-page clustering \cite{strehl:cluster00}. To do a K-way partitioning, the multi-level graph partitioning algorithm first collapses the graph into a sequence of increasingly smaller graphs. The spectral partitioning algorithm is then used to partition the smallest graph. The resulting partitions are then projected back to the original graph through series of transformations. 

After a topic is partitioned into multiple subtopic partitions, a topic from each partition is selected as its partition label. A good label for a partition should describe most of the topics contained within the partition. Therefore, we pick the topic with the highest total connection to other topics in the same cluster to be the partition label. Such a topic is the most centralized node within that partition. Alternatively, other strategies such as selecting the topic with the highest strength can also be used. Our results, however, do not find them to be as effective. 

\section{Evaluation}

Evaluating the quality of an automatically generated concept hierarchy is challenging. One possible method is to compare it against another existing manually created hierarchy or gold standard. Precision and Recall then can be used to measure the accuracy and the completeness of the discovered hierarchy. However, a gold standard evaluation has two drawbacks. First, it assumes that there exists one true optimal hierarchy, which is generally not the case. The second and more practical concern is that there might not even be a suitable manually created hierarchy to compare against. The alternative is to have humans manually judge the quality of the generated hierarchies. This approach allows for the possibility of multiple coherent hierarchies. However, such human-based evaluation is time consuming and subjective. In this work, we evaluated our GraBTax algorithm using both human-based and gold standard approaches. In the first experiment, human assessors were asked to compare the hierarchies generated by GraBTax with those produced by hierarchical agglomerative clustering. In the second experiment, GraBTax results were compared against synthetic gold standard created from Wikipedia.

\subsection{Human-based Evaluation}

In the human-based evaluation, two metrics are used to assess the quality of a taxonomy: Precision (P) and Semantic Precision (SP). We did not calculate Recall since it requires all candidate topics to be individually examined for each query. 

\textbf{\textit{Precision}} measures the percentage of topics in a taxonomy that are relevant to the root (the query topic) and is defined as the number of relevant topics divided by the total number of topics in the taxonomy. Topics that are errors from the automatic extraction are always counted as irrelevant. 

\textbf{\textit{Semantic Precision}} measures the quality of the relations in the taxonomy and is the percentage of relevant topics in the taxonomy that are classified under their semantically relevant parents. More precisely, SP is equal to the number of relevant topics that are under their correct parents divided by the total number of topics in the taxonomy. SP of a taxonomy is always lower or equal to P. Only relevant topics are counted in the calculation of SP. A topic that is classified under its correct super-topic is a semantically relevant topic. For example, in Figure \ref{fig:mt}, ``\textit{query translation}'' is semantically relevant under ``\textit{cross language information retrieval},'' while ``\textit{nlp system}'' is not semantically relevant under ``\textit{word sense disambiguation}.''

\subsubsection{Experiment Results}

We apply GraBTax to 1.1M papers (with titles and abstracts) from CiteSeer$^X$. The 376,577 keyphrases were extracted as the topic set. For evaluation, six well-known topics in computer science were selected to be the query topics: ``\textit{artificial intelligence},'' ``\textit{information retrieval},'' ``\textit{machine translation},'' ``\textit{computer graphics},'' ``\textit{semantic web},'' and ``\textit{social networks}.'' Since to manually evaluate the full taxonomy for each query topic would take too much time (for example, the fully expand taxonomy for ``\textit{information retrieval}'' contains more than 1,000 topics), we restrict the maximum number of topics in each taxonomy to be 200. The threshold $r_{max}$, $k_{min}$ and $s_{min}$ in the subgraph selection step are set as 3, 10, 20 respectively. The exact number of topics included for each query topic is  shown in Table \ref{table:prec}. All 6 taxonomies contain up to 5 levels of subtopics. 
The top two-level of the hierarchy for ``\textit{machine translation}'' is shown in Figure \ref{fig:mt}.

Three graduate students in computer science were asked to manually assess the relevancy and the semantic relevancy of each topic in all 6 taxonomies produced by GraBTax. Their assessments are then averaged and are used to compute Precision and Semantic Precision for each taxonomy. In addition, we also make a comparison with a hierarchal agglomerative clustering method (HAC) as the baseline approach. For comparison, for each query, HAC is given the same set of topics as the one generated by GraBTax to cluster. Each topic is represented with a document vector indicating document set that the topic appears in. The Pearson correlation coefficient is used as the distance function. The distance between two topic clusters are computed using the centroid-linkage. Since HAC generally does not generate labels for clusters, for the purpose of evaluating the semantic relevancy we asked the human assessors to use the following strategy to assign a label for each cluster. If a cluster is merged with another singleton cluster, then use the singleton cluster as the label for the merged cluster. If both clusters have more than one topic, then the assessors are instructed to select the best of the two cluster labels for the merged cluster. Given the types of relationships we are interested in, e.g. ``\textit{bag of words model}'' as a sub-concept of ``\textit{natural language processing}'' (as in Wikipedia category hierarchy), we did not compare our results with standard pattern-based approaches such as \cite{hearst:hyponym92}, since most such relationships are not found in the corpus with the typical is-a, part-of patterns. 

Precision and Semantic Precision for both GraBTax and HAC are shown in Table \ref{table:prec}. Since HAC uses the set of topics generated by GraBTax as the input, the Precision values are the same as GraBTax's. In general, the Precision values are quite high for all queries, implying that most topics in the generated taxonomies are relevant to the query topic. ``\textit{machine translation}'' has the highest Precision at 0.97, while ``\textit{artificial intelligence}'' has the lowest at 0.77. GraBTax is better at discovering meaningful relationship compared to HAC. Its average Semantic Precision is 0.69, which is significantly higher than that of HAC. HAC seems to be able to cluster highly related topics very well, but does poorly otherwise. HAC's taxonomies also tend to be unbalanced, compared to that of GraBTax's. The micro average agreements between 3 human judges are 0.63 for P and 0.8 for SP. Their average Cohen's Kappa is 0.35 for P, which is fair, and 0.66 for SP, which is quite good.

\begin{table} \footnotesize
\centering	
\begin{tabular}{|l|r|l|l|l|l|} \hline
\multirow{2}{*}{Query Topic}	& \multirow{2}{*}{\#Topics} & \multicolumn{2}{|c|}{GraBTax} & HAC \\ \cline{3-5}
								&		   & P & SP & SP \\ 
\hline \hline
computer graphics		& 200		& 0.88	& 0.68	& 0.42 \\ \hline
information retrieval	& 200		& 0.90	& 0.72	& 0.48 \\ \hline
artificial intelligence	& 185		& 0.77	& 0.63	& 0.46 \\ \hline
semantic web			& 161		& 0.86	& 0.68	& 0.40 \\ \hline
machine translation		& 128 		& 0.97	& 0.71	& 0.44 \\ \hline
social network			& 82		& 0.88 	& 0.69 	& 0.45 \\ \hline
\hline
Average					& 159.33 	& 0.88	& 0.69	& 0.44 \\ \hline
\end{tabular}
\caption{Precision (P) and Semantic Precision (SP) for each of the six query topics\label{table:prec}}
\end{table}

\subsection{Wikipedia Evaluation}

While human-based evaluations are appropriate for assessing quality of subjective items such as concept taxonomies, they are time consuming to produce. In contrast, evaluating an automatically generated result against a gold standard can be done quickly and cheaply. In addition to being more objective, a gold standard comparison can be used for parameter tuning. As such we set out to construct a gold standard.

Unfortunately, existing taxonomies for concepts in computer science such as ODP categories and the ACM Classification System\footnote{http://www.acm.org/about/class/ccs98-html} are unsuitable as a gold standard. ODP categories are too broad and do not contain the majority of concepts produced by our algorithm. For instance, there are no sub-concepts for \textit{``Semantic Web''} in ODP. Also some portions of ODP categories under computer science are not computer science related concepts, especially at the lower level. For example, the concepts under \textit{``Neural Networks''} are \textit{Books}, \textit{People}, \textit{Companies}, \textit{Publications}, \textit{FAQs, Help and Tutorials}, etc. The ACM Classification System has similar drawbacks, where its categories are too broad for comparison.

Thus, we instead opted to construct the gold standard from Wikipedia's categories and page titles, which better intersect with our topic set. Unlike ODP, Wikipedia also contains fine-grain concepts such as \textit{brill tagger}, and \textit{chart parser}. Six gold standard taxonomies were built for six categories under Computer Science in Wikipedia: ``\textit{artificial intelligence},'' ``\textit{human-computer interaction},'' ``\textit{software engineering},'' ``\textit{natural language processing},'' ``\textit{programming languages}'' and ``\textit{computer graphics}.'' 

To construct a gold standard taxonomy for a query category, we first built the taxonomy tree containing all Wikipedia categories and page titles under that query category up to the depth of four (not counting the root level). We limited the depth because, given that Wikipedia categories are not a taxonomy by design, following the subcategory links too deeply can result in a leaf concept that is irrelevant to the query category (e.g. ``\textit{artificial intelligence $\Rightarrow$ search algorithms $\Rightarrow$ internet search algorithms $\Rightarrow$ URL normalization}''). Then we prune away all the concepts that are not in a candidate set. Additionally, every concept that does not have at least one candidate topic as its ancestor is also filtered out. For example, the category ``\textit{artificial intelligence $\Rightarrow$ history of artificial intelligence $\Rightarrow$ Herbert Simon}'' is not included in the gold standard because ``\textit{history of artificial intelligence}'' is not in our topic set. The final total number of concepts for each six categories are shown in Table \ref{table:wiki}.

For each of the six topic queries, we used the GraBTax algorithm to generate a taxonomy from the topic-specific subgraph that contains all the categories in that query's gold standard taxonomy. We then computed the accuracy of parent-child relationships found in the generated taxonomy with respect to the gold standard. This experiment specifically evaluates the last step of the GraBTax algorithm (Section \ref{sec:partition}), which is how well it partitions a topic-specific subgraph into a taxonomy (not on the precision of term selection). Two scoring metrics are used to determine whether a concept is correctly placed in a taxonomy: exact match and partial match. 
\begin{table*}\footnotesize
\centering	
\begin{tabular}{|l|r|r|r|r|r|r|r|} \hline
Query Topic				& \# Rels & $Exact$ & $Partial$ & $Exact_{\lambda_2=0}$ & $Partial_{\lambda_2=0}$ \\
\hline \hline
artificial intelligence		& 379		& 17.2\% & 31.2\% & 14.8\%	& 24.9\% \\ \hline
software engineering		& 613		& 7.7\%	 & 17.7\% & 7.2\%	& 15.1\% \\ \hline
human-computer interaction	& 216		& 4.2\%	 & 24.0\% & 6.9\%	& 16.5\% \\ \hline

computer graphics			& 292		& 3.4\%	 & 16.0\% & 4.1\%	& 15.8\% \\ \hline
natural language processing	& 198 		& 8.5\%	 & 23.1\% & 13.1\%	& 20.1\% \\ \hline
programming languages		& 33		& 24.2\% & 32.6\% & 24.2\%	& 32.6\% \\ \hline
\hline
Average						& 288.5 	& 10.9\%& 24.1\% & 11.7\%& 20.8\% \\ \hline
\end{tabular}
\caption{matches of the generated taxonomies compared against the gold standards from Wikipedia in percentage\label{table:wiki}}
\end{table*}

\textbf{\textit{Exact Match}}. A concept is considered an exact match if and only if its parent in the gold standard is also its immediate parent in the taxonomy. 

\textbf{\textit{Partial Match}}. A concept is considered to be a partial match if its paths to the root node in the gold standard and in the taxonomy share a common intermediate concept. Thus, for partial match, even if its parent is incorrect, a concept is still given a partial credit if its path to the root resembles the correct path. More specifically, for a concept $C$, where a concept $A$ is its nearest common ancestor, its score is computed as $\frac{1}{p \times q}$. $p$ is the distance between $C$ and $A$ in the taxonomy and $p$ is the distance between $C$ and $A$ in the gold standard. If such a concept $A$ does not exists for $C$, then the score is $0$. To illustrate, for a concept $C$, where its path in the gold standard is ``$R$ $\Rightarrow$ $A$ $\Rightarrow$ $C$'' and its path in the taxonomy is ``$R$ $\Rightarrow$ $A$ $\Rightarrow$ $B$ $\Rightarrow$ $C$'' ($R$ is the root concept), $C$ is not an exact match, since its parent in the taxonomy is $B$, not $A$. But $C$ is a partial match, since both paths share a common non-terminal concept $A$ and its partial match score is $\frac{1}{1 \times 2} = 0.5$. The partial score will be $\frac{1}{1 \times 1} = 1$ for an exact match case. 

\subsubsection{Experiment Results}

Table \ref{table:wiki}. shows the exact match scores and the partial match scores for all six categories when compared with the gold standards. 
For exact match, 11\% of concepts are placed under their correct parents. For partial match, the average matching score is 24\%. This is reasonable considered that if a node is inserted into a perfect taxonomy between a root node and all its children, the partial match score will already be 50\%. One reason for the difficulty is that many concepts can be validly placed under multiple paths. For instance, the path for ``\textit{SAS system}'' in Wikipedia is 
``\textit{NLP $\Rightarrow$ information retrieval $\Rightarrow$ data management $\Rightarrow$ business intelligence},'' while GraBTax puts it under ``\textit{NLP $\Rightarrow$ data mining $\Rightarrow$ data warehouse},'' resulting in no match. We also feel that term selection given by Wikipedia contribute to some of errors, e.g. ``\textit{SAS system}'' is reachable from ``\textit{NLP}'' in Wikipedia even though they are not related. This results in many irrelevant nodes needed to be placed in the taxonomy. 

The two $[\lambda_2=0]$ runs illustrate the improvement contributed by introducing lexical similarity to the graph model. The overall partial match improves from 21.5\% to 25.1\%. Interestingly, the exact match score shows slight drop in performance. We think this is because of the difficulty in producing the exact match leading to high variance in performance. 

\section{Conclusion And Future Work} \label{sec:length}

We propose a graph-based algorithm for taxonomy generation, GraBTax and apply our algorithm to build taxonomies for topics in computer science. 
Through user studies, we show that our algorithm generates a taxonomy containing relevant subtopics, and is superior to a hierarchal clustering approach for discovering semantic relations. In addition, we propose a method using a gold standard to empirically evaluate the performance of the algorithm against Wikipedia categories, which can be used for parameter tuning. Our experimental results show that the generated taxonomies are in reasonable agreement with Wikipedia categories. For the future, we plan to improve the quality of generated taxonomies by introducing a post-generation refinement step. We intend to explore multiple refinement strategies using various topological features, such as the clustering coefficient, sibling connectivity, and parent-child connectivity. In addition, we plan to evaluate the effectiveness of alternative strategies in subgraph selection and the cluster label selection. 


\bibliographystyle{naaclhlt2012}
\bibliography{hlt12-puck}

\end{document}